\documentclass[
  aip,jcp,
  reprint 
]{revtex4-1}

\usepackage{amsmath}
\usepackage{amsfonts}
\usepackage{booktabs}
\usepackage{pbox}
\usepackage{url}
\usepackage{graphicx}
\usepackage{color}
\usepackage{soul,xcolor} 
\usepackage{cancel}

\newcommand{\dd}[2]{\frac{d#1}{d#2}}

\newcommand{\bi}{\mathbf{i}}
\newcommand{\bj}{\mathbf{j}}
\newcommand{\bzero}{\mathbf{0}}
\newcommand{\bH}{\mathbf{H}}

\newcommand{\bI}{\mathbf{I}}

\newcommand{\bk}{\mathbf{k}}
\newcommand{\br}{\mathbf{r}}

\newcommand{\bR}{\mathbf{R}}
\newcommand{\bD}{\mathbf{D}}

\newcommand{\bU}{\mathbf{U}}

\newcommand{\bra}{\langle}
\newcommand{\ket}{\rangle}
\newcommand{\sgn}{\mathrm{sgn}}

\newcommand{\eref}[1]{Eq.~(\ref{#1})}

\newcommand{\fref}[1]{Fig.~\ref{#1}}
\newcommand{\tref}[1]{Table~\ref{#1}}
\newcommand{\aref}[1]{Appendix~\ref{#1}}

\begin{document}

\setstcolor{red}

\title{The effect of quantization on the FCIQMC sign problem}

\author{M.H.~Kolodrubetz}
\affiliation{Department of Physics, Princeton University, Princeton, NJ 08544, U.S.A.}
\affiliation{Department of Physics, Boston University, Boston, MA 02215, U.S.A.}
\author{J.S.~Spencer}
\affiliation{Department of Materials, Imperial College London, Exhibition Road, London SW7 2AZ, U.K.}
\affiliation{Department of Physics, Imperial College London, Exhibition Road, London SW7 2AZ, U.K.}
\author{B.K.~Clark}
\affiliation{Department of Physics, Princeton University, Princeton, NJ 08544, U.S.A.}
\affiliation{Princeton Center for Theoretical Science, Princeton University, Princeton, NJ 08544, U.S.A.}
\affiliation{Station Q, Microsoft Research, Santa Barbara, CA 93106, U.S.A.}
\author{W.M.C.~Foulkes}
\affiliation{Department of Physics, Imperial College London, Exhibition Road, London SW7 2AZ, U.K.}

\begin{abstract}
  The sign problem in Full Configuration Interaction Quantum Monte Carlo 
  (FCIQMC)
  without annihilation 
  can be understood as an instability of the psi-particle
  population to the ground state of the matrix obtained by making all
  off-diagonal elements of the Hamiltonian negative.  Such a matrix, and
  hence the sign problem, is basis dependent.  In this paper we discuss
  the properties of a physically important basis choice: first versus
  second quantization.  For a given choice of single-particle
  orbitals, we identify the conditions under which the
  fermion sign problem in the second quantized basis of 
antisymmetric Slater determinants
is identical to the sign problem in the first 
  quantized basis of 
  unsymmetrized Hartree products.  We 
also show that, when the two
  differ, the fermion sign problem is always less severe 
  in the second quantized basis.
  This supports the idea that FCIQMC, even in the absence of
  annihilation, improves
  the sign problem relative to first quantized methods.  Finally, 
we point out some theoretically interesting classes of 
Hamiltonians where first and second quantized sign problems
differ, and others where they do not.
\end{abstract}

\maketitle

\section{Introduction}

Projector quantum Monte Carlo methods such as Diffusion Quantum Monte
Carlo\cite{Foulkes2001_1} (DMC) and Full Configuration Interaction
Quantum Monte Carlo\cite{Booth2009_1} (FCIQMC) generate a stochastic
representation of the solution of the imaginary-time Schr\"{o}dinger
equation, $\partial |\psi\ket/\partial \tau = -\widehat{H}|\psi\ket$. As long as
the starting state $|\psi(\tau=0)\ket$ has a non-zero overlap with the
ground state $|\psi_0\ket$, the solution $|\psi(\tau)\ket$
converges to $|\psi_0\ket$ as $\tau \rightarrow \infty$,
up to a normalization.
In FCIQMC, the Hamiltonian
is represented in a finite discrete basis, typically Slater
determinants.  The imaginary-time Schr\"{o}dinger equation is thus
expressed in the matrix formulation,
\begin{equation}
  \dd{c_\bi(\tau)}{\tau} = - \sum_{\bj} \bra \bi | \left( \widehat{H}
  - S \widehat{I} \right) | \bj \ket c_\bj(\tau), 
  \label{eqn:itSE}
\end{equation}
where $|\bi\ket$ represents a many-particle basis function (e.g. a
determinant), $c_\bi(\tau)=\bra \bi | \psi(\tau)\ket$ is the component
of $|\psi(\tau)\ket$ along $|\bi\ket$, $S$ is a parameter which can be
adjusted to control the normalization, and $\widehat{I}$ is the identity
operator. We define $H_{\bi\bj} \equiv \bra \bi | ( \widehat{H} -
S\widehat{I} ) | \bj \ket$ and assume that $\widehat{H}$ has
time-reversal symmetry, so that $H_{\bi \bj}$ and thus $c_{\bi}(\tau)$
can be chosen to be real.

In FCIQMC, an initial population of signed \emph{psi-particles} or
\emph{psips} (not to be
confused with the real particles of the system) is distributed over the
Hilbert space such that the expected value of the signed psip population
on $|\bi\ket$ is proportional to $c_{\bi}(0)$. The psip distribution is
then evolved in time using an algorithm which ensures that the expected
value of the signed population on $|\bi\ket$ remains proportional to
$c_{\bi}(\tau)$. In the $\tau\rightarrow\infty$ limit, the psip
distribution provides a stochastic snapshot of the ground state
wave function.\cite{Foulkes2001_1,Booth2009_1,Spencer2012_1}

The amplitudes $c_{\bi}(\tau)$ and $c_{\bi}(\tau + \Delta\tau)$ of a
solution to \eref{eqn:itSE} are related 
by the formula
\begin{equation}
c_{\bi}(\tau + \Delta\tau) = 
\sum_{\bj} \langle \bi | e^{-(\widehat{H} - S\widehat{I})\Delta\tau} |
\bj \, \rangle c_{\bj}(\tau),
\label{eqn:timestep}
\end{equation}
for which different projector 
QMC methods use different approximations for the
exponential. 
For instance, DMC in the continuum uses a Trotter
approximation to generate the transition matrix 
$\bU_\textrm{DMC}({\bf
  R}, {\bf R}') \equiv \langle {\bf R} | e^{-\widehat T \Delta \tau}
 e^{-(\widehat{V}-S\widehat{I}) \Delta \tau} |{\bf
  R}'\rangle$, where $\widehat{T}$ and $\widehat{V}$ are the
kinetic and potential energy 
operators, and $\bR$ and $\bR'$
are positions in the  $3N$-dimensional space of 
particle coordinates.  In
FCIQMC, a first-order finite difference approximation is used instead:
$e^{-\bH\Delta\tau} \approx \bI - \bH \Delta\tau \equiv
\bU_\textrm{FCIQMC}$.  
If the product of the time step $\Delta \tau$ and the
largest eigenvalue $E_\textrm{max}$ of $H$ is not too large, then
the ground state of $\bH$
is the same as the \emph{dominant} eigenvector of 
$\bU_\textrm{FCIQMC}$. Note that throughout this
paper we use the
term `dominant'  to mean the eigenvector with the
largest eigenvalue.
Similarly, in Lattice Regularized DMC (LRDMC), 
where an upper bound on $\widehat H$ is introduced by 
adding an artificial lattice\cite{Casula2005_1}, one also uses
the transition matrix ${\bf U}_\textrm{LRDMC} =
{\bf I} - {\bf H}\Delta\tau$.  
In this paper, we focus primarily on the case of FCIQMC.  However, by
formulating our arguments in terms of the transition matrix $\bU$,
most of our results are applicable to all projector QMC methods.

\section{General properties of the sign problem}

FCIQMC is a stochastic algorithm for
applying the real symmetric matrix $\bU_{\textrm{FCIQMC}}$ (which 
we call $\bU$ from now on) to a population of psips.  
Because FCIQMC works in a discrete basis, two psips
of opposite sign occasionally end up on the same basis function, at
which point they cancel out and can be removed from the simulation.
When FCIQMC was initially proposed, it was observed
that this process, known as `annihilation,'
 is critical in order for the psip
distribution to converge to the ground state\cite{Booth2009_1}.
Two of us subsequently
showed\cite{Spencer2012_1} that:
\begin{enumerate}
\item In the absence of annihilation, the densities of positive
  ($\{c_\bi^+\}$) and negative ($\{c_\bi^-\}$) psips evolve according to
  coupled equations.  The out-of-phase component, $c_\bi^+
  - c_\bi^-$, evolves according to 
  \begin{eqnarray}
    \lefteqn{c_{\bi}^+(\tau + \Delta\tau) - c_{\bi}^-(\tau +
      \Delta\tau)} \notag \\
    & \hspace*{2em} = & 
    \sum_{\bj} {U}_{\bi\bj} \left [ c_{\bi}^+(\tau) - c_{\bi}^-(\tau)
    \right ]
  \end{eqnarray}
  and converges to the dominant eigenstate of $\bU$, i.e., the
  ground state of $\bH$.  In contrast, the evolution of the in-phase
  component, $c_\bi^+ + c_\bi^-$, is governed by a different matrix
  $\widetilde{\bU}$, the elements of which are the absolute values of
  the elements of $\bU$:
\begin{equation}
\widetilde U_{\bi \bj} = \big| U_{\bi \bj} \big| ~.
\end{equation}

\item The largest eigenvalue of $\widetilde{\bU}$ is \emph{always}
  greater than or equal to the largest eigenvalue of $\bU$. Thus, in the
  absence of annihilation, both $c_{\bi}^+$ and $c_{\bi}^-$ tend to the
  dominant eigenvector of $\widetilde{\bU}$. The physical ground state can in principle be
  obtained by subtracting $c_{\bi}^-$ from $c_{\bi}^+$, but the
  difference is exponentially smaller than $c_{\bi}^+$ and
  $c_{\bi}^-$. In a real simulation using a finite number of psips, the
  difference is swamped by the statistical noise in $c_{\bi}^+$ and
  $c_{\bi}^-$; this is the sign problem in the FCIQMC method. The
  severity of the sign problem depends on the difference between the
  largest eigenvalues of $\bU$ and $\widetilde{\bU}$.
\end{enumerate}

For projector methods such as FCIQMC and LRDMC, which are based on a
transition matrix of the form $\bU \equiv \bI - \bH \Delta
\tau$, we can talk instead about the properties of $\widetilde{\bH}$,
the matrix obtained by making all off-diagonal elements of $\bH$
negative:
\begin{equation}
\widetilde H_{\bi \bj} = - \big| H_{\bi \bj} \big|\, ; \;\;\; 
\widetilde H_{\bi \bi} = H_{\bi \bi} ~.
\end{equation}
Since the diagonal elements of $\bU$ are always positive for small
enough $\Delta\tau$, the diagonals of $\widetilde \bU$ and $\bU$
match. By construction, so do the diagonals of $\widetilde \bH$ and
$\bH$.  Thus, for small enough $\Delta \tau$, $\widetilde \bU =
\bI - \widetilde \bH \Delta \tau$, so the ground state of $\widetilde
\bH$ is identical to the dominant eigenstate of $\widetilde \bU$.

\section{Comparison of first and second quantized sign problems}

The analysis of the sign problem in Ref.~\onlinecite{Spencer2012_1} did
not assume anything about the choice of basis, and in particular did not
require that the basis be first quantized (unsymmetrized, as in DMC
simulations \cite{Foulkes2001_1,Assaraf2007_1,Casula2005_1}) or second
quantized (symmetrized or antisymmetrized, as in FCIQMC simulations of
bosonic or fermionic systems, respectively\cite{Booth2009_1}).  In this
paper, we compare the sign problems for many-fermion systems expressed
in first and second quantized bases, answering two questions.  First,
when are the sign problems of first and second quantized algorithms
different?  Second, when they differ, which is better?

In order to be clear about the basis in which a quantity is expressed,
we shall use the subscripts $F$, $D$ and $P$ to indicate a first
quantized basis of Hartree products, a second quantized basis of Slater
determinants, and a second quantized basis of permanents, respectively.
A quantity without any subscript is general and can be considered in any
of the three bases.

Let us consider a basis of Hartree products, $\{|\bi\ket\}$, for a
system of $N$ fermions and $M$ mutually orthogonal single-particle basis functions, where
$|\bi\ket = |\phi_{i_1}(1) \phi_{i_2}(2) \cdots
\phi_{i_N}(N)\ket$ and multiple occupancy of any single-particle basis
function, $\phi_j$, is forbidden.  The vector index $\bi$ defines the
list of $N$ single-particle basis functions appearing in a Hartree
product and specifies the order in which those basis functions occur.
The first quantized Hamiltonian is formed by taking matrix elements of
the Hamiltonian operator with respect to the $^{M}P_N$ Hartree products,
so that each eigenstate of the Hamiltonian can be written as a linear
combination of the form $|\psi\ket = \sum_\bi c_\bi |\bi\ket$.
However, physically meaningful many-fermion
wave functions must be totally antisymmetric with respect to exchange of
any two fermions and hence may also be expressed as linear combinations
of Slater determinants.

The fermionic second quantized basis is thus the set of $^{M}C_N$
determinants $\{ |D_\bi\ket \}$.  We use the notation $|\bi\ket \in
|D_\bi\ket$ to indicate that the Hartree product $|\bi\ket$ 
is a permutation of the orbitals in $|D_\bi\ket$.
For $|\bi\ket \in |D_\bi\ket$,  $|D_\bi\ket$ can be
obtained from $|\bi\ket$ using the antisymmetrization operator
$\mathcal A$, defined by
\begin{equation}
\begin{split}
&\mathcal A | \phi_{i_1} (1) \phi_{i_2} (2) \cdots \phi_{i_N} (N) \ket = \\
&\frac{1}{N!}\sum_{P} (-1)^{\zeta_P} |\phi_{i_{P(1)}} (1) \cdots 
\phi_{i_{P(N)}} (N) \ket ~,
\end{split}
\end{equation}
where $P$ labels permutations of the $N$ orbitals and $\zeta_P$ is 
the sign of the permutation.  Then 
  $\mathcal A |\bi\ket = \frac{\mathrm{sgn}\langle\bi|D_{\bi}\rangle}{\sqrt{N!}}
  |D_\bi\ket,$
where the factor of $\sqrt{N!}$ ensures that $|\bi\ket$ and 
$|D_\bi\ket$ remain normalized.  The order of the orbitals appearing in a determinant
$|D_\bi\ket = (N!)^{1/2} \mathcal{A} |\phi_{i_1}(1) \phi_{i_2}(2) \cdots
\phi_{i_N}(N)\ket$ affects only the overall sign. Hence, to uniquely
specify each determinant, we insist that $i_1 < i_2 < \cdots < i_N$.  We
utilize a similar notation for the (bosonic) basis of permanents, $\{
|P_\bi\ket \}$, defined by
$|P_\bi\ket = \sqrt{N!}\, \mathcal S |\bi\ket$, where 
$\mathcal S$ is the symmetrizer. 
Throughout this paper, we exploit the fact that there is a 
one-to-one mapping between $|D_\bi\ket$ and $|P_\bi\ket$,
which holds because the single-particle orbitals are 
orthogonal and multiple occupancy of any given single-particle orbital is
forbidden.  In the context of the bosonic permanents, this corresponds
to an effective hardcore constraint in the space of single-particle orbitals.

\begin{figure}
  \begin{center}
  \includegraphics[width=\linewidth]{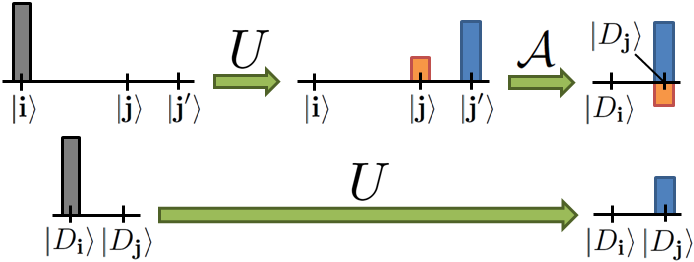}
  \end{center}
  \caption{A single step of the FCIQMC algorithm, illustrating how the
    first and second quantized sign problems can differ.  (Top) 
In the first
    quantized algorithm, a psip on $|\bi\ket$ 
may spawn children on two
    different Hartree products, $|\bj\ket$ and $|\bj'\ket$, belonging to
    the same determinant: $|\bj\ket, |\bj'\ket \in |D_\bj\ket$.  After
    antisymmetrizing, the psips may contribute with opposite sign to
    the total weight on $|D_\bj\ket$.  (Bottom) 
Performing the same step in the
    second quantized algorithm, the cancellation on $|D_\bj\ket$ is
    automatically accounted for by the sign and magnitude of $\bra D_\bi
    | \widehat U | D_\bj \ket$.}
  \label{fig:psips}
\end{figure}

To get a feel for how the sign problem could be worse in a first
quantized formalism than in second quantized one,
consider a first quantized psip 
on $|\bi\ket$ 
that is connected by off-diagonal elements of $\widehat U$
to $|\bj\ket$ and $|\bj'\ket$. If the orbitals in
$\bj$ and $\bj'$ are permutations of
one another, then $|\bj\ket$ and $|\bj'\ket$ both appear in the
expansion of the same determinant: $|\bj\ket,|\bj'\ket \in |D_\bj\ket$.
After a single step of the FCIQMC algorithm, it is possible for the
antisymmetrizer to map these two psips to $|D_\bj\ket$ with opposite
signs, as illustrated in \fref{fig:psips}. 
If the amplitude of $|\psi(\tau)\ket$ on $|D_{\bj}\ket$ is small,
the cancellation of the two first quantized contributions needs to be
very accurate to stochastically obtain the correct amplitude. 
In practice, the small signal is swamped by statistical
fluctuations in the first quantized psip populations and a sign problem
results.  FCIQMC in a determinant basis performs
this cancellation automatically.  Therefore, as we rigorously
prove in \aref{appendix:proof}, the first and
second quantized sign problems are
identical if and only if the two
first quantized psips always contribute with the same sign to the total
weight on $|D_\bj\ket$. This argument also suggests that, if the
sign problems differ, the second quantized algorithm will be better.  

To define the difference between sign problems 
more explicitly, we consider the ground
state properties of $\widetilde \bU$ in a first and second quantized basis.
In the first quantized basis, both $\bU_F$ and $\widetilde \bU_F$
commute
with arbitrary permutations of the particle
labels.  Therefore, the dominant 
eigenstate $|\widetilde \varphi_F\ket$ of
$\widetilde \bU_F$ must transform according to an irreducible
representation of the permutation group.  Since all elements of
$\widetilde \bU_F$ are non-negative in the $|\bi\rangle$ basis, it may
be shown via a simple variational argument that all non-zero components
$\langle \bi |\widetilde \varphi_F\ket$  have the same sign, and hence that
$|\widetilde \varphi_F\ket$ has a non-zero overlap with the bosonic
state $|\psi_B\ket=\sum_\bi |\bi\ket$.  As eigenvectors transforming according to
different irreducible representations are orthogonal, it follows that $|\widetilde
\varphi_F\ket$ must also be a bosonic state. 
Therefore, $|\widetilde
\varphi_F\ket$  may equally well be regarded as the
dominant eigenvector of the matrix $\bra P_\bi | \widetilde \bU_F |
P_\bj \ket$ obtained by evaluating the matrix elements of $\widetilde
\bU_F$ in the basis of permanents.

Meanwhile, the second quantized 
algorithm is unstable with
respect to the dominant eigenstate of the matrix $\widetilde \bU_D$ with
elements $|\langle D_\bi | \bU_F | D_\bj \rangle|$.  Although the
matrices $\langle P_\bi | \widetilde \bU_F | P_\bj \rangle$ and
$\widetilde \bU_D$ have the same size -- each containing $(N!)^2$ fewer
elements than $\widetilde \bU_F$ in a Hartree product basis -- they are
not in general the same matrix.  Whenever they differ,  we prove in 
\aref{appendix:proof} that the dominant eigenvalue
of $\widetilde \bU_D$ is smaller than the dominant eigenvalue of
$\langle P_i | \widetilde \bU_F | P_j \rangle$; the second quantized
sign problem is therefore less severe than the first quantized sign
problem. The sign problems in the first and second quantized bases are
the same if and only if the two matrices are the same:
\begin{equation}
  \bra P_\bi | \widetilde  \bU_F | P_\bj \ket
  = \bra D_\bi | \widetilde \bU_D | D_\bj \ket .
  \label{eq:perm_det_equiv}
\end{equation}
In the Appendix, we also prove that this condition is equivalent to 
our previous conditions on the psip  dynamics, as discussed above
and illustrated in \fref{fig:psips}.

It is worth noting that
in FCIQMC and LRDMC, where $\widetilde \bU_F = \bI - \widetilde \bH_F
\Delta \tau$, $|\widetilde \varphi_F \rangle$ is also the bosonic ground
state of $\widetilde \bH_F$.  This state is not in general the same as
the physical many-boson ground state, which is the lowest-energy 
totally-symmetric eigenstate of $\bH_F$. 
For LRDMC in real
space, however, all off-diagonal elements of $\bU_F$ are already
positive, so $\bU_F = \widetilde \bU_F$, 
$\bH_F = \widetilde \bH_F$, and 
the instability \emph{is}  with respect to the physical bosonic ground
state for this special case.

\begin{figure} 
  \begin{center}
  \includegraphics[width=0.8\linewidth]{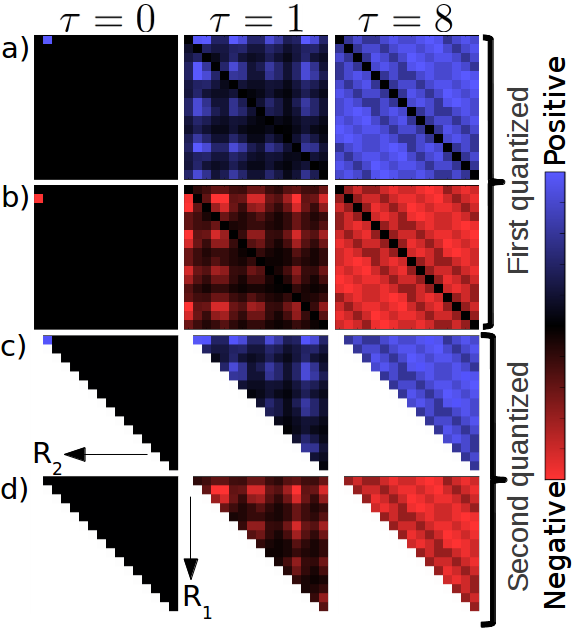}
  \end{center}
  \caption{Comparison of first and second quantized psip distributions
    for an FCIQMC simulation of two spinless 
fermions interacting via the Hamiltonian $\widehat
    H=-\sum_{\langle i j\rangle} \left[ c_i^\dagger c_j + c_j^\dagger
      c_i - n_i n_j\right]$, where the sum is over nearest neighbors of
    a  $4\times4$ square lattice with periodic boundary
    conditions.  Images show the probability distributions of positive
    (a,c) and negative (b,d) psips separately as functions of the
    imaginary time $\tau$. The propagation is performed analytically
    according to \eref{eqn:timestep}
using the 
approximation $\bU_\mathrm{FCIQMC}=\bI-\bH \Delta \tau$ 
with time step $\Delta\tau=0.1$.  While the
    positive and negative psip populations individually converge to the
    ground state of $\widetilde \bH$ ($E=-7.41589748$), their
    exponentially small difference (not shown) converges to the ground
    state of $\bH$ ($E=-5.89171753$).\cite{toy_fci_code} Note
    that the upper right-hand triangles of comparable distributions
    (positive first quantized and positive second quantized, for
    example) match.  The ``missing'' triangles can be reconstructed by
    expanding out the Slater determinants, after which the psip
    distributions are step-by-step identical.  
    A psip located at $(R_i,R_j)$
    represents the Hartree product  $|\phi_i(1)
    \phi_j(2)\ket$ or Slater determinant $\sqrt{2} \mathcal{A}|\phi_i(1) \phi_j(2)\ket$,
    where $\phi_i$ is a single-particle basis function at the lattice
    vertex $(x_i, y_i)$ and $R_i=x_i+4y_i$.  }
  \label{fig:snapshots}
\end{figure}

\section{Examples for physical Hamiltonians}

There are a number of physical situations in which the equivalence
condition, \eref{eq:perm_det_equiv}, 
holds. The simplest is when the
two-particle part of the Hamiltonian operator is diagonal in the
Hartree product basis, so that all off-diagonal matrix elements are
single-particle in nature.
Important such physical examples are the Hubbard model in real space and
the LRDMC algorithm. For these Hamiltonians, the psips accumulate a
(diagonal) weight according to the electronic potential and move
(off-diagonally) according to the one-body kinetic operator. 
To illustrate the relationship between first and second quantized psip
populations in such a system, \fref{fig:snapshots} plots their
distributions for a simple problem involving two spinless fermions
hopping between nearest-neighbor sites on a two-dimensional square
lattice.  As expected, the first and 
second quantized psip distributions are
step-by-step identical, after antisymmetrization.

A more general case where the sign problems are equivalent 
is when, in the space of determinants, all exchange 
integrals in diagonal matrix elements are non-negative
and all off-diagonal matrix elements involve only a
single integral.
To understand these conditions, consider an arbitrary
Hamiltonian with one-particle interaction $\hat h$ and 
two-particle interaction $\hat u_{12}$.  If only one
off-diagonal integral is non-zero, then
off-diagonal elements on both sides of \eref{eq:perm_det_equiv} are just the negative
absolute value of the non-zero term\cite{Footnote_DiagonalTerms}.
Similarly, the restriction on diagonal exchange integrals is
necessary to make the diagonal elements of 
\eref{eq:perm_det_equiv} match. As exchange terms occur only in
the off-diagonal elements of the first quantized Hamiltonian $\bH_F$, they
enter $\bra P | \widetilde \bH_F | P \ket$ with negative magnitude
by the definition of $\widetilde \bH_F$:
\begin{align}
\bra P | \widetilde \bH_F | P  \ket =& \frac{1}{N!} \sum_{\bi \in P} 
    \bra \bi | H_F |\bi  \ket + \frac{1}{N!} \sum_{\bi \in P} \sum_{\substack{\bj \in P  \\ 
           \bj \neq \bi \hspace{0.03in}}}
\Big[ - \big| \bra \bi | H_F | \bj  \ket \big| \Big] \nonumber \\
=& \sum_i \bra i | \hat h | i  \ket + \sum_{i<j} \bra ij|\hat u_{12}|ij \ket -
\sum_{i<j} \big| \bra ij|\hat u_{12}|ji \ket \big|~,
\label{eq:PHfP}
\end{align}
where $\phi_{i,j}$ are single-particle basis
functions occupied in $D$ and $P$. 
Meanwhile, $\bra  D | \widetilde \bH_D | D  \ket$ is simply
the matrix element of the second quantized Hamiltonian:
\begin{align}
\bra  D | \widetilde \bH_D | D  \ket =& \bra  D | \bH_D | D  \ket \nonumber \\
=& \sum_i \bra i | \hat h | i  \ket + \sum_{i<j} \bra ij|\hat u_{12}|ij \ket -
\sum_{i<j} \bra ij|\hat u_{12}|ji \ket ~.
\label{eq:DHdD}
\end{align}
Clearly Eqs.~(\ref{eq:PHfP}) and ~(\ref{eq:DHdD}) match if and only if the 
diagonal exchange integrals are positive.

The Hubbard model in momentum space is an example where these conditions
are met:
First, there are no exchange contributions to the diagonal
Hamiltonian matrix elements.  Second, every member $|\bj\ket$ of the set $\{|\bj\ket\}$ of
Hartree products obtained by applying the
momentum-space Hubbard Hamiltonian to an arbitrary Hartree product
$|\bi\ket$ differs from $|\bi\ket$ 
by at most one spin-up orbital
and one spin-down orbital.  As no two such Hartree products in $\{|\bj\ket\}$ can 
contain the same set of occupied
orbitals, no two Hartree products in $\{|\bj\ket\}$
can appear in the expansion of the same determinant.

In the generic case, however, the first and second
quantized sign problems are not identical.  For instance, the matrix
element of $\widetilde \bH_D$ between two determinants $D$ and
$D_{ab}^{ij}$, where $D_{ab}^{ij}$ is obtained from $D$ by replacing the
single-particle orbitals $\phi_a$ and $\phi_b$ by $\phi_i$ and $\phi_j$
without re-ordering, is \cite{Slater1929_1,Condon1930_1,Szabo1998_1}
\begin{equation} 
  - \Big| \bra ij | \hat{u}_{12} | ab \ket - 
  \bra ij | \hat{u}_{12} | ba \ket \Big|,
\end{equation}
where $\hat{u}_{12}$ is the two-particle interaction 
operator, e.g., the Coulomb interaction
$\hat{u}_{12} = 1/|\br_1 - \br_2|$. The equivalent element of
$\widetilde \bH_F$ in a permanent basis is
\begin{equation}
  - \Big| \bra ij | \hat{u}_{12} | ab \ket \Big| 
  - \Big| \bra ij | \hat{u}_{12} | ba \ket \Big|.
\end{equation}
In general, these two matrix elements will not be equivalent. 
As a concrete example, consider a gas of four electrons at a density of
$r_s = 1\;\text{a.u.}$ subject to periodic boundary
conditions. \tref{table:UEG} shows the ground state energy eigenvalues
of the matrices obtained when this system is studied using a basis of
(antisymmetrized) product functions constructed from a set of 38
one-electron plane waves. The sign problem in a basis of determinants is
less severe than that in a basis of Hartree products.

\begin{table}
  \begin{ruledtabular}
    \begin{tabular}{ll}
      $M_{\bi\bj}$ & Lowest eigenvalue (a.u.) \\
      \hline
      $(\bH_F)_{\bi\bj} = \bra\bi|\widehat{H}|\bj\ket$ &  5.63133019 \\
      $(\widetilde \bH_F)_{\bi\bj} = -|\bra\bi|\widehat{H}|\bj\ket|$ & 5.26282782  \\
      $(\bH_D)_{D_\bi D_\bj} = \bra D_\bi|\widehat{H}| D_\bj\ket$ 
      &  5.63133019 \\
      $(\widetilde \bH_D)_{D_\bi D_\bj} = -|\bra D_\bi|
      \widehat{H}|D_\bj\ket|$ & 5.34900281 \\
      $(\widetilde \bH_F)_{P_\bi P_\bj}$ \\
      $\quad = -\frac{1}{4!}\sum_{\substack{|\bi\ket\in|P_\bi\ket \\ 
          |\bj\ket\in|P_\bj\ket}} | \bra\bi| \widehat{H} |\bj\ket |$
      & 5.26282782 
    \end{tabular}
  \end{ruledtabular}
  \caption[UEG eigenvalues]{Lowest eigenvalues of various matrices
    $M_{\bi\bj}$ related to the sign problem in the 3D uniform electron
    gas.\cite{toy_fci_code}  Periodic boundary conditions were applied
    to a simulation cell of dimensions $L\times L \times L$ containing
    four electrons at a density of $r_s=1\ \text{a.u.}$  The basis set
    consisted of all 38 plane waves with momentum less than $2\pi/L$.
    The resulting determinant and permanent Hilbert spaces each contain
    567 functions with $M_s=0$ and momentum $\bk = \bzero$.  The
    corresponding Hartree-product Hilbert space contains 13608 basis  
    functions.
\label{table:UEG}
}
\end{table}

\section{Conclusions}

In summary, we have given two equivalent conditions for determining whether the
first and second quantized sign problems for a given system are the same. For
situations where the sign problems differ, we have shown that the second
quantized algorithm is better. Thus, the use of an explicitly antisymmetrized
basis can ameliorate the most fundamental problem faced by almost all fermion
QMC methods: the sign problem. Indeed, our conditions show that this
improvement \emph{will} occur for nearly all problems involving QMC on
realistic systems (atoms, molecules, solids, etc.). However, we have also seen
that there are a number of lattice models of great interest to the condensed
matter community, such as the Hubbard model, where second quantization has no
effect on the sign problem.


There remain many open questions regarding the sign problem in FCIQMC.  
It has now been shown that both annihilation and second quantization 
improve the sign problem, but their relative effectiveness remains 
uncertain in general.  Another interesting question is how the improved 
sign problem in FCIQMC compares to the fermion Monte Carlo (FMC) method 
introduced by Kalos and Pederiva\cite{Kalos2000_1}, which uses correlated walks to enhance 
annihilation.  In this latter method, Assaraf et al.\cite{Assaraf2007_1} showed that  the 
cancellation step similarly modifies the eigenvalue toward which FMC is unstable.  
Based on the results of these papers, one might then expect that combining correlated 
walks with second quantization in certain systems
(e.g., the UEG in momentum space) might further 
attenuate the effects of the fermion sign problem.  

Finally, we note that although our arguments and proofs have been focused on
the case of fermions, they can be readily extended to many-boson systems and
used to show that working in a second quantized basis of permanents is at least
as good as working in a first quantized basis of Hartree products.

\appendix
\section{Proof of equivalence conditions} \label{appendix:proof}

We will now show: a) that the conditions described in the main text
uniquely characterize when the first and second quantized sign problems
differ, and b) that, if they differ, the second quantized sign problem
is less severe.  We work with matrices formed from the operator
$\widehat U$ introduced in the main text, 
such that the wave functions sampled at time steps $n$
and $n+1$ are related by $|\psi_{n+1}\ket = \widehat U | \psi_n\ket$.  We
continue to use the notation $\widetilde \bU_\alpha$ to denote a matrix whose
elements are the absolute values of those in $\bU_\alpha$, i.e.,
\begin{eqnarray}
\bra \bi | \widetilde \bU_F | \bj \ket &=& 
\big| \bra \bi | \widehat U | \bj \ket \big| \nonumber \\
\bra D_\bi | \widetilde \bU_D | D_\bj \ket &=& 
\big| \bra D_\bi | \widehat U | D_\bj \ket \big| \nonumber \\
\bra P_\bi | \widetilde \bU_P | P_\bj \ket &=& \big| 
\bra P_\bi | \widehat U | P_\bj \ket \big| ~.
\label{eq:Utilde_def}
\end{eqnarray}

We have already shown in the main text that {\bf if 
\eref{eq:perm_det_equiv} holds, then the first and second quantized sign problems are
identical.} The remainder of this appendix will proceed to prove the following statements:
\begin{itemize}
\item The conditions in \eref{eq:perm_det_equiv} can be restated in terms of 
first quantized matrix elements (\eref{eq:equiv_conds}).
\begin{itemize}
\item Therefore, if \eref{eq:equiv_conds} holds, then the first and second 
quantized sign problems are identical.
\end{itemize}
\item If the conditions in \eref{eq:equiv_conds} do not hold, then the 
second quantized sign problem is strictly better than the first quantized sign problem.
\begin{itemize}
\item Therefore, the first and second quantized 
sign problems are identical if and only if \eref{eq:equiv_conds} holds.
\end{itemize}
\end{itemize}

\subsection{Restatement of conditions for equivalence}

As noted in
\eref{eq:perm_det_equiv} from the main text, it is clear 
that the first and second quantized FCIQMC algorithms
have the same sign problem if
\begin{equation}
\label{eq:fpos_eq_dpos}
  \bra P_\bi | \widetilde \bU_F  | P_\bj \ket = 
  \bra D_\bi |  \widetilde \bU_D | D_\bj \ket ~.
\end{equation}
Noting that the permanent $|P_\bj \ket$ and determinant $| D_\bj \ket$
occupy the same set of orbitals, we can expand them as
\begin{equation}
\label{eq:perm_det_exp}
|P_\bj\ket = \frac{1}{\sqrt{N!}} \sum_{\bj \in D_\bj} | \bj \ket ~, ~~~
|D_\bj\ket = \frac{1}{\sqrt{N!}} \sum_{\bj \in D_\bj} | \bj \ket ~
\mathrm{sgn} \bra \bj | D_\bj  \ket ~.
\end{equation}
Then, using Eqs.~(\ref{eq:Utilde_def}) and~(\ref{eq:perm_det_exp}),
\eref{eq:fpos_eq_dpos} can be rewritten
\begin{eqnarray}
 \bra D_\bi |  \widetilde \bU_D | D_\bj \ket  &=& \bra P_\bi | \widetilde \bU_F  | P_\bj \ket
   \nonumber \\
& \Updownarrow & \nonumber \\
\Big | \bra D_\bi | \widehat U | D_\bj \ket \Big | &=& 
\frac{1}{N!} \sum_{\substack{\bi \in D_\bi\\\bj \in D_\bj}}
\bra \bi | \widetilde \bU_F | \bj \ket \nonumber \\ 
& \Updownarrow & \nonumber \\
\Bigg| \sum_{\substack{\bi \in D_\bi\\\bj \in D_\bj}} 
  \bra \bi | \widehat U | \bj \ket  
  \mathrm{sgn}\bra \bi | D_\bi \ket
  \mathrm{sgn}\bra \bj | D_\bj \ket \Bigg|
 &=& 
\sum_{\substack{\bi \in D_\bi\\\bj \in D_\bj}} 
\left| \bra \bi | \widehat U | \bj \ket \right| ~.
\label{eq:pre_triangle}
\end{eqnarray}
By the triangle inequality, the right-hand side 
of \eref{eq:pre_triangle} is greater than or equal
to the left-hand side, with equality if and only if all terms within
the absolute values on the left-hand side are of the same sign,
$s_{D_\bi D_\bj}$.  Since equality does hold, it
follows that, for all $\bi \in D_\bi$ and $\bj \in D_\bj$, then either
\begin{equation}
\begin{aligned}
  &\bullet \; \bra \bi | \widehat{U} | \bj \ket = 0, ~~\text{or}\\
  &\bullet \; \mathrm{sgn}\bra \bi | \widehat{U} |
  \bj \ket \mathrm{sgn} \bra \bi| D_\bi \ket \mathrm{sgn}\bra \bj |
  D_{\bj} \ket = s_{D_{\bi} D_{\bj}}  , &&
\end{aligned}
\end{equation}
where the sign $s_{D_{\bi} D_{\bj}}$ is the same for all $\bi \in
D_{\bi}$ and $\bj \in D_{\bj}$.  If we consider two distinct FQ basis
elements $\bj, \bj' \in D_\bj$, we deduce that either
\begin{equation}
\begin{aligned}
  & \bullet \;
  \bra \bi | \widehat U | \bj \ket = 0 ~~\text{or}\\
  & \bullet \;
  \bra \bi | \widehat U | \bj' \ket = 0 ~~\text{or}\\
  & \bullet \; \sgn \bra \bj | D_\bj \ket \sgn \bra \bi | \widehat U |
  \bj \ket = \sgn \bra \bj' | D_\bj \ket \sgn \bra \bi | \widehat U |
  \bj' \ket ~.
\end{aligned}
\label{eq:equiv_conds}
\end{equation}
Note that these are precisely the conditions we assumed were met by the first quantized
psip distribution in the main text (cf. \fref{fig:psips}).

We have now shown that {\bf the conditions in \eref{eq:perm_det_equiv} and
\eref{eq:equiv_conds} are equivalent}. As a corollary, {\bf if 
the conditions in \eref{eq:equiv_conds} hold, then the first and
second quantized sign problems are identical.}

\subsection{If \eref{eq:equiv_conds} is false, then second quantized 
is better than first quantized}

We now prove that,
if the conditions in \eref{eq:equiv_conds} are
not met, then the sign problem in a second quantized basis is less severe
than the sign problem in a first quantized basis. 
Therefore, for the remainder of this section, assume that
\eref{eq:equiv_conds} does not hold.

Let $|\widetilde
\varphi_D\ket$ be the eigenstate of $\widetilde \bU_D$ with the largest
eigenvalue $\widetilde t_D$.  Construct the (normalized) FQ state
$|\varphi_F\ket$ with components
\begin{equation}
  \bra \bi | \varphi_F \ket = \frac{1}{\sqrt{N!}} 
  \bra D_\bi | \widetilde \varphi_D \ket ~,
\end{equation}
where $|\bi\ket \in |D_\bi\ket$.  Since $\widetilde \bU_D$ is a
non-negative real symmetric matrix, all amplitudes $\bra D_{\bi} |
\widetilde \varphi_D \ket$ (and thus $\bra \bi | \varphi_F \ket$) 
have the same sign and may be chosen real and
non-negative.

Now consider the expectation value $\widetilde{t}_F'$ of the matrix
$\widetilde{\bU}_F$ in the state $|\varphi_F \ket$:
\begin{equation}
  \widetilde t_F' \equiv \bra \varphi_F | \widetilde \bU_F | \varphi_F \ket
  = \sum_{\bi,\bj} \bra \varphi_F | \bi \ket 
  \bra \bi | \widetilde \bU_F | \bj \ket
  \bra \bj | \varphi_F \ket ~.
\end{equation}
From \eref{eq:Utilde_def},
$\bra \bi | \widetilde \bU_F | \bj \ket = 
\left| \bra \bi | \widehat U | \bj \ket \right|$.  Hence
\begin{align}
  \nonumber
  \widetilde t_F' &= \frac{1}{N!} \sum_{\bi,\bj} \bra D_\bi | \widetilde \varphi_D \ket
  \bra D_\bj | \widetilde \varphi_D \ket \left| \bra \bi | \widehat U | \bj \ket \right|\\
  &= \frac{1}{N!} \sum_{D_\bi, D_\bj} \bra D_\bi | \widetilde \varphi_D \ket
  \bra D_\bj | \widetilde \varphi_D \ket \sum_{\bi \in D_\bi}\sum_{\bj \in D_\bj}
  \left| \bra \bi | \widehat U | \bj \ket \right| ~.
  \label{eq:det_sect}
\end{align}
In the last step, we have explicitly broken
the sums over $\bi$ and $\bj$ into
sums over determinants $D_{\bi}$ and $D_{\bj}$ and sums over the Hartree
products contained in those determinants:
\begin{equation}
\sum_\bi = \sum_{D_\bi} \sum_{\bi \in D_\bi} ~.
\end{equation}

Since, by assumption, the conditions in
\eref{eq:equiv_conds} are not met, for at least one $D_\bj$ there must exist some $\bj,
\bj' \in D_\bj$ such that
\begin{equation}
\begin{aligned}
  & \bullet \; 
  \bra \bi | \widehat U | \bj \ket \neq 0 \mbox{\, and}\\
  & \bullet \;
  \bra \bi | \widehat U | \bj' \ket \neq 0 \mbox{\, and}\\
  & \bullet \;
  \sgn \bra \bj | D_\bj \ket \sgn \bra \bi | \widehat U | \bj \ket \neq
  \sgn \bra \bj' | D_\bj \ket \sgn \bra \bi | \widehat U | \bj' \ket ~.
\end{aligned}
\label{eq:no_equiv_conds}
\end{equation}
Consider the term
\begin{equation}
\sum_{\bj \in D_{\bj}}  \left| \bra \bi | \widehat U | \bj \ket \right|
\end{equation}
appearing on the right-hand side of \eref{eq:det_sect}. 
Since $\bra \bj | D_\bj \ket \neq 0$, this summation is equal to
\begin{equation}
\sum_{\bj \in D_{\bj}}  \left| \bra \bi | \widehat U | \bj \ket
  \mathrm{sgn} \bra \bj | D_{\bj} \ket \right|.
\end{equation}
\eref{eq:no_equiv_conds} tells us that the signs of the $\bj$ and $\bj'$
contributions to this summation differ before the absolute value
is taken. Hence, by the triangle inequality,
\begin{equation*}
\sum_{\bj \in D_{\bj}}  \left| \bra \bi | \widehat U | \bj \ket
  \mathrm{sgn} \bra \bj | D_{\bj} \ket \right| > \left| \sum_{\bj \in D_{\bj}} \bra \bi | \widehat U | \bj \ket
  \mathrm{sgn} \bra \bj | D_{\bj} \ket \right| ~,
\end{equation*}
and therefore, substituting back into \eref{eq:det_sect},
\begin{equation}
  \widetilde t_F' > \frac{1}{N!} \sum_{D_\bi, D_\bj} \bra D_\bi | \widetilde \varphi_D \ket
  \bra D_\bj | \widetilde \varphi_D \ket \sum_{\bi \in D_\bi} \left| \sum_{\bj \in D_\bj}
  \bra \bi | \widehat U | \bj \ket \sgn \bra \bj | D_\bj \ket \right|.
\label{eq:triangle}
\end{equation}
Noting that $[\widehat H, \mathcal A]$ (which implies
that $[\widehat U, \mathcal A]=0$) and that
\begin{equation}
\sum_{\bj \in D_\bj} | \bj \ket \sgn \bra \bj | 
D_\bj \ket = \sqrt{N!} |D_\bj \ket ~,
\end{equation}
we have
\begin{eqnarray}
\nonumber
\Big| \sum_{\bj \in D_\bj} \bra \bi | \widehat U | \bj \ket \sgn \bra \bj | D_\bj \ket \Big| & = &
\Big| \sqrt{N!} \bra \bi | \widehat U | D_\bj \ket \Big|\\
\nonumber
&=& \Big| \sqrt{N!} \bra \bi | \widehat U \mathcal A | D_\bj \ket \Big|\\
\nonumber
&=& \Big| \sqrt{N!} \bra \bi | \mathcal A \widehat U | D_\bj \ket \Big|\\
\nonumber
&=& \Big| \bra D_\bi | \widehat U | D_\bj \ket \sgn \bra \bi | D_\bi \ket \Big| \\ 
&=& \Big| \bra D_\bi | \widehat U | D_\bj \ket \Big| ~. 
\end{eqnarray}
We may now rewrite \eref{eq:triangle} in terms of determinants
\begin{equation}
\widetilde{t}_F' > \frac{1}{N!} \sum_{D_\bi,D_\bj} \bra D_\bi |
\widetilde{\varphi}_D \ket \bra D_\bj | \widetilde{\varphi}_D \ket
\sum_{\bi \in D_{\bi}}  \left|
  \bra D_\bi | \widehat U | D_\bj \ket  \right|.
\end{equation}
The sum over ${\bi \in D_{\bi}}$ cancels the $1/N!$ prefactor to give
\begin{eqnarray}
  \nonumber
  \widetilde t_F' &>& \sum_{D_\bi D_\bj} \bra D_\bi | \widetilde \varphi_D \ket
  \bra D_\bj | \widetilde \varphi_D \ket \left| 
  \bra D_\bi | \widehat U | D_\bj \ket \right|\\
  &=& \sum_{D_\bi D_\bj} \bra D_\bi | \widetilde \varphi_D \ket
  \bra D_\bj | \widetilde \varphi_D \ket 
  \bra D_\bi | \widetilde \bU_D | D_\bj \ket = \widetilde t_D .
\end{eqnarray}
Therefore, by the variational principle, the largest eigenvalue
$\widetilde{t}_F$ of $\widetilde{\bU}_F$ is greater than the largest
eigenvalue $\widetilde{t}_D$ of $\widetilde{\bU}_D$, meaning
that {\bf if \eref{eq:equiv_conds} does not hold, then 
the first quantized sign problem is strictly worse than the
second quantized sign problem}.

\subsection{Conclusions}

In conclusion, we have shown that $\widetilde t_F = \widetilde t_D$ if
and only if the conditions in \eref{eq:equiv_conds} hold.  If they do
not hold, then $\widetilde t_F > \widetilde t_D$.  The second quantized
sign problem is therefore less severe than the first quantized sign
problem unless the conditions in \eref{eq:equiv_conds} are satisfied.
Finally, we reiterate that, as long as the largest eigenvalues of $\bH$
and $\widetilde \bH$ are finite, their ground states are the same as the
dominant eigenstates of $\bU=\bI - \bH \Delta \tau$ and $\widetilde \bU
= \bI - \widetilde \bH \Delta\tau$, respectively, in the limit 
$\Delta \tau \rightarrow 0$.  Therefore, for FCIQMC and LRDMC, the
statements in this appendix can be recast in terms of ground state
properties of the Hamiltonian matrix.

\bibliography{quant}

\begin{thebibliography}{11}
\bibitem [{Foo()}]{Footnote_DiagonalTerms}%
  \BibitemOpen
  \href@noop {} {}\bibinfo {note} {Consider, for example, the case where
  $\bD_\bi$ and $\bD_\bj$ differ by a single orbital, such that $\phi_i$ is an
  orbital occupied in $\bD_\bi$, $\phi_j$ is in $\bD_\bj$, and $\phi_a$ is in
  both determinants. If only one integral is non-zero, e.g., the exchange
  integral $\bra i a | \hat u_{12} | a j \ket$, then the off-diagonal terms of
  \eref{eq:perm_det_equiv} are given by $\bra D_\bi | \widetilde \bH_D | D_\bj
  \ket = \bra P_\bi | \widetilde \bH_F | P_\bj \ket = - \big| \bra i a | \hat
  u_{12} | a j \ket \big|$.}\BibitemShut {Stop}%
\end{thebibliography}

\end{document}